\documentclass[USenglish]{article}	

\usepackage[utf8]{inputenc}				
\usepackage{lmodern} 
\usepackage{microtype}
\usepackage[numbers,square,sort&compress]{natbib}
\usepackage{caption}
\usepackage{subcaption}
\usepackage{enumitem}
\usepackage{booktabs} 
\usepackage{graphicx}
\usepackage[margin=1in]{geometry}

\usepackage{amsmath}
\usepackage[utf8]{inputenc}
\DeclareMathOperator\erf{erf}

\usepackage[toc,title,page]{appendix}
\usepackage{multirow}
\usepackage{hyperref}

\begin{document}

\author{Zach Rosenof}
\title{Dynamic quantification of player value for fantasy basketball}
	

\maketitle

\includegraphics[scale = 0.7]{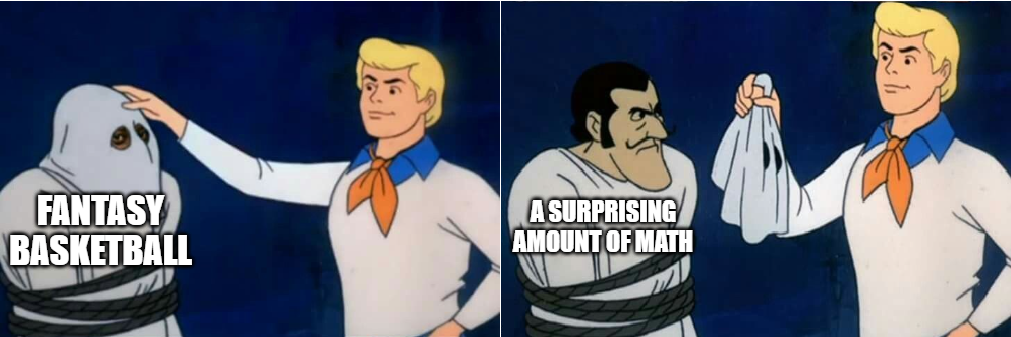}
	
\abstract{Previous work on fantasy basketball quantifies player value for category leagues without taking draft circumstances into account. Quantifying value in this way is convenient, but inherently limited as a strategy, because it precludes the possibility of dynamic adaptation. This work introduces a framework for dynamic algorithms, dubbed ``H-scoring'', and describes an implementation of the framework for head-to-head formats, dubbed $H_0$. $H_0$ models many of the main aspects of category league strategy including category weighting, positional assignments, and format-specific objectives. Head-to-head simulations provide evidence that $H_0$ outperforms static ranking lists. Category-level results from the simulations reveal that one component of $H_0$'s strategy is punting a subset of categories, which it learns to do implicitly.}

\section{Introduction}

See previous work for an introduction to fantasy basketball and definitions relevant to its mathematical study (Rosenof, 2024). 

Much of that previous work has been on static ranking systems and how they can be used to rank players. However, static ranking systems are inherently limited because they do not allow for adaptive strategies. 

``Punting'' is one such adaptive strategy. When a manager punts, they strategically sacrifice one or more categories in order to improve performance in the rest. The choice of which categories to punt is necessarily informed by the players a manager has already chosen and how strong they are in each category. So the strategy is inherently adaptive, and cannot be executed properly with a static ranking list. 

This work has two purposes. One is to introduce a novel framework for a dynamic algorithm with adapts to drafting circumstances, called H-scoring. The other is to describe and analyze an implementation of the framework called $H_0$

\section{H-scoring}

The central premise of H-scoring is that the aggregate statistics of future draft picks can be approximated as a heuristic function of one or more parameters encoding the manager's strategy. Those parameters can be optimized for each player, allowing the manager to select the player associated with the highest value player-parameter set.

The framework requires three pre-defined functions
\begin{itemize}
\item $X(j)$ estimates probability distributions for the differences in category totals between the manager's team and opposing teams. $j$ is a set of parameters which encodes the manager's strategy for future picks 
\item $W(X(j))$ or $W(j)$ calculates category victory probabilities based on the distributions estimated via $X(j)$. It is equivalent to the CDF of the overall category differential distribution at zero 
\item $V(W(X(j)))$ or $V(j)$ defines the objective function, tailored for the scoring format
\end{itemize}

The framework is then simply 

\begin{enumerate}
\item Estimate an optimal $j$ to maximize $V(j)$ for each player, with gradient descent or some other optimization algorithm
\item Choose the player with the highest $V(j)$
\end{enumerate}

This framework is designed with snake drafts in mind. For a discussion of how to apply it to value players for auctions, see Appendix \ref{AuctionConv}

\section{The $H_0$ algorithm}

$H_0$ is an implementation of H-scoring. Its parameter set $j$ encodes two levers of influence over future draft picks- how to weigh different categories against each other, and which positions to prioritize. $X(j)$, $W(j)$ and $V(j)$ are defined such that $V(j)$ is differentiable, and it optimizes $j$ via gradient descent

\subsection{Assumptions}

$H_0$ makes use of many simplifying assumptions

\begin{itemize}

\item The manager's team must fit into a pre-defined positional structure. So long as it does, all games played by all chosen players count 

\item Player performance distributions are known exactly, and do not change over the course of a season. There are no substitutions or trades

\item All players' performances have standard deviation $m_\tau$ for counting statistics and $r_\tau$ for percentage statistics

\item The distribution of performance means across all available players, relative to future picks that will be made by opposing teams, can be approximated with a particular form. More detail on this assumption is provided in Section \ref{CategoryWeights}

\item Percentage statistics can be treated equivalently to counting statistics, once in the basis of a static ranking system

\item Category-level performances are independent for each player. Therefore, category victory probabilities are also independent

\item The manager's goal is to maximize their expected performance over an arbitrary scoring period. The definition of performance varies based on format 

    \begin{itemize}
    \item Each Category: number of categories won against an arbitrary opponent
    \item Most Categories: one if winning a match-up against an arbitrary opponent, otherwise zero
    \end{itemize}    

\item If the manager in question has chosen $K$ players, $K+1$ players are known for all other teams. In the case when $N$ opponents have only selected $K$ players, averages of the next $N$ players in G-score order are used to fill in the $K+1$ player slots 

\item For the purpose of calculating variance, it can be assumed that

\begin{itemize}
    \item There is no variance in the aggregate statistics of the manager's future draft picks
    \item The variances of other managers' draft picks are equivalent to what they would be if those managers were choosing players at random
\end{itemize}

\item Any locally optimal $j$ is sufficient

\end{itemize}

$H_0$ aims to optimize the fantasy basketball problem based on these assumptions. How well the assumptions reflect actual fantasy basketball is discussed in Section \ref{Ass}

\subsection{Modeling category differences $X$} \label{X}

Formulating $X(j)$ requires a model of expected statistics for relevant players. For this purpose, it is helpful to define a new scoring system called X-score, which is the G-score with the $m_{\sigma}$ and $r_{\sigma}$ terms set to zero. For counting statistics, that is 

$$
X_p = \frac{m_p - m_{\mu}}{m_{\tau}}
$$

And for percentage statistics

$$
X_p = \frac{\frac{a_q}{a_{\mu}} \left( r_q - r_{\mu} \right) }{r_{\tau}}
$$

Like in the static context, these need some estimate of $Q$, the set of relevant fantasy players. 

X-scores are helpful because while week-to-week variance merits roughly the same treatment in the dynamic context as it does in the static, player-to-player variance must be treated differently. Managers know the average statistics of previously picked players, rendering those players' effective player-to-player variance zero.

Another helpful quantity to define is $v$, which is a vector that converts from the X-score basis to the G-score basis. For counting statistics, it is 

$$
\frac{m_{\tau} }{\sqrt{m_{\tau}^2 + m_{\sigma}^2}}
$$

And for percentage statistics it is

$$
\frac{r_{\tau} }{\sqrt{r_{\tau}^2 + r_{\sigma}^2}}
$$

For convenience, $v$ is normalized to sum to one

\subsubsection{Team decomposition}

Considering the match-up against team $O$, a total of $2N$ players are relevant. As depicted by Table \ref{fig:H-score} these $2N$ players can be broken down into five groups: 

\begin{itemize}
\item $q \in A_c$, already chosen players on team $A$. Size $= K$
\item $w = p$, the single player to be chosen by the manager of team $A$ with their current pick 
\item $q \in A_u$ unknown remaining team A players. Size $= N - K - 1$
\item $q \in O_m$, known players on team O. Size $= K + 1$
\item $q \in O_u$, unknown players on team O, matching up to unknown players on team $A$. Size $= N - K - 1$
\end{itemize}

\begin{table}[ht]
\centering
\begin{tabular}{| l | c | c |}
\hline
& Team $A$ & Team $O$ \\
\hline
Pick 1 & \multirow{5}{*}{$A_c$} & \multirow{6}{*}{$O_m$} \\
Pick 2 & & \\
Pick 3 & & \\
Pick 4 & & \\
Pick 5 & & \\
\cline{1-2} 
Pick 6 & p &\\[1pt]
\hline
Pick 7 & \multirow{7}{*}{$A_u$} & \multirow{7}{*}{$O_u$}\\
Pick 8 & & \\
Pick 9 & & \\
Pick 10 & &\\
Pick 11 & & \\
Pick 12 & & \\
Pick 13 & & \\
\hline
\end{tabular}

\caption{Relevant player sets, $K = 5$ and $N = 13$}
\label{fig:H-score}
\end{table}

Applying the central limit theorem, $H_0$ describes the differential between two teams in the basis of X-scores as follows

$$
X(j) = \mathcal{N} \left(X_s + X_p + X_{\delta} - X_{O_m} , 2N + (N-K-1) X_{\sigma}^2 \right) 
$$

Where 
\begin{itemize}
\item $X_s$ is the aggregate expected performance of players already selected on team $A$, in terms of X-score
\item $X_p$ is the X-score of the candidate player 
\item $X_\delta$ is a measure of how different the mean performances of picks in $A_u$ are expected to be from those of the picks in $O_u$, in the X-score basis. In the H-scoring framework it depends on $j$, so it can also be written as $X_\delta(j)$
\item $X_{O_m}$ is the aggregate expected performance of players already selected by the opponent
\item $X_{\sigma}^2$ is player to player variance, again in the basis of X-scores. It is estimated as the variance of X-scores over $Q$
\end{itemize}

The mean is a decomposed description of team $O$'s expected performance minus team $A$'s expected performance. The variance has $2N$ because in the X-score basis, $m_\tau$ and $r_\tau$ are cancelled out by the X-score denominator, leaving a standard variance of one for each player. The other contribution to variance is from the players in $O_u$, under the assumption that they are chosen at random.

Most of the components of the distribution are simple to calculate. The one complicated component is $X_\delta(j)$, which is a function of how the manager plans to behave in the future and must be approximated.

This decomposition is designed with a snake draft in mind. An alternative for auctions is presented in Appendix \ref{AuctionDecomp}

\subsubsection{Adjusting for category weights } \label{CategoryWeights}
 
$H_0$ models category prioritization strategy with a weight vector $j_C$, which always sums to one. $H_0$ imagines that the manager would use that weight vector for the rest of the draft, and approximates $X_\delta$ based on that. For example, if $j_C$ was set to $v$, then the manager would be using $v$-weighted X-scores for their future picks, which are equivalent to G-scores. $H_0$ would then expect future picks to be balanced. But if it was different, say 

$$
j_C = \begin{bmatrix}
\frac{1}{8} & \frac{1}{8} & \frac{1}{8} & \frac{1}{8} & \frac{1}{8} & \frac{1}{8} & \frac{1}{8} & \frac{1}{8} & 0 \\
\end{bmatrix}
$$

Then $H_0$ would imagine that the manager would behave differently. In this case, it would expect the ninth category to be punted, leading to a skewed distribution across categories. 

The math on what to expect from future picks is complicated and requires liberal use of approximations about the space of available players and their expected performances. Details are in Appendix \ref{apdx.xdel}. More briefly, it is assumed that player statistics are distributed as a multivariate normal distribution, and opposing managers make their remaining picks based on G-scores or equivalently $v$-weighted X-scores. The standard deviation of the manager's weighting of a category difference off baseline for an arbitrary candidate player $j_C^T x_{\delta_q}$, dubbed $\sigma$, roughly determines how different the chosen player can be expected to be from a generic player. It is estimated that in the $j_C$ basis the chosen player is $\omega \sigma$ stronger, and in the $v$ basis they are $\gamma \sigma$ weaker. Conditionally, the expected player that maximizes $j_C^T x_{\delta_q}$ then has a difference off a generic player of 

$$
X_\delta(j_C) = \left(N-K-1\right) * \Sigma * \left( v j_C^T - j_C v^T \right) * \Sigma * \frac{
   \left( - \gamma j_C - \omega v \right) \sqrt{\left(j_C -  \frac{v v^T \Sigma j_C}{v^T \Sigma v} \right) ^T \Sigma \left( j_C -  \frac{v v^T \Sigma j_C}{v^T \Sigma v}  \right) }
  }{j_C^T \Sigma j_C * v^T \Sigma v - \left( v^T \Sigma j_C \right) ^2}
$$

Where $\Sigma$ is the covariance matrix between categories for player performance means (calculated separately across positions and averaged). $\gamma$ and $\omega$ can be estimated empirically, by observing actual values when the algorithm is run and fitting them with a linear regression.

Deriving the derivative with respect to $j_C$ is a matter of extensive routine calculation. It can be calculated unless $j_C = v$, in which case $X_\delta(j_C)$ is undefined and therefore cannot be differentiated

\subsubsection{Adjusting for positions} \label{Positions}

One of the assumptions of $H_0$ is that all teams must fit a certain positional structure. An example structure is

\begin{itemize}
\item Three utility spots (any position)
\item Two centers
\item Two guards (point guard or shooting guard)
\item One point guard
\item One shooting guard
\item Two forwards (power forward or small forward)
\item One power forward
\item One small forward 
\end{itemize}

There are potentially many options for assigning already drafted players to slots, both since there are flexible slot types like utilities, and because players are often eligible for multiple positions. $H_0$ models this as an assignment problem, and solves the problem outside of the gradient descent framework.

An assignment problem is a problem where $N$ nodes need to be connected one-to-one with another $N$ nodes, with each potential connection being associated with a reward. The goal is to maximize the total reward. $H_0$ uses the following reward structure for its assignment problem

\begin{itemize}
\item There is no reward for assigning players already drafted to desirable slots, since their statistics are locked in and do not change based on how they are categorized. Their rewards are zero for slots that they are eligible for and $- \infty$ for those which they are not eligible. \item Future players can be any slot, but different slots may have different rewards depending on what kind of statistics the manager is looking for. Positional rewards for the non-flex slots are estimated as $\mu_C j_C$, where $\mu_C$ is a matrix of average category values per position (weighed by number of eligible positions, e.g. if a player is eligible for four positions they have $25\%$ weight for each one)
\item Flex slot rewards for future players are the highest rewards among eligible positions. E.g. the reward for guard is the highest between the reward for shooting guard and point guard. There is also a small flex bonus, to ensure that flex slots will be prioritized for future players- .0001 for guard/forward, and .0002 for utility 
\end{itemize}

Based on this reward structure, $H_0$ solves for the optimal assignments and uses that to infer how many players it will take of each slot type with future picks. Because of the way the reward structure is designed, $H_0$ tries to place already drafted players at less desirable slots, freeing up desirable slots for future players.  

Say that a manager already has a center $p_0$, and is considering drafting a player $p_1$ who is eligible as either a center or power forward. Their $j_C$ does not weight assists highly, and therefore their position rewards are skewed against assist-heavy point guard as such:

$$
\begin{bmatrix}
C = 0.5 & PG = -0.4 & SG = -0.2 & PF = 0.4 & SF = 0.3 
\end{bmatrix}
$$

Then their assignment matrix would be Table \ref{Tab:PosAss}. In this example, the manager's highest-value option is to place $p_0$ at center and $p_1$ at power forward, because those are the least valuable slots possible to assign them to. The manager then has room for three utilities, one center, two guards, one of each guard type, two forwards, and one small forward for their remaining picks.
\begin{table}[!h]
\centering
\begin{tabular}{l r r r r r r r r r r r r r}
 & Utility1 & Utility2 & Utility3 & C1 & C2 & G1 & G2 & PG & SG & F1 & F2 & PF & SF \\ \midrule
$p_0$ & 0 & 0 & 0 & 0 & 0 & $- \infty$ & $- \infty$ & $- \infty$ & $- \infty$ & $- \infty$ & $- \infty$ & $- \infty$ & $- \infty$\\
$p_1$ & 0 & 0 & 0 & 0 & 0 & $- \infty$ & $- \infty$ & $- \infty$ & $- \infty$ & 0 & 0 & 0 & $- \infty$\\
$p_2$ & 0.5002 & 0.5002 & 0.5002 & 0.5 & 0.5 & -1.999 & -1.999 & -0.4 & -0.2 & 0.4001 & 0.4001 & 0.4 & 0.3 \\
$p_3$ & 0.5002 & 0.5002 & 0.5002 & 0.5 & 0.5 & -1.999 & -1.999 & -0.4 & -0.2 & 0.4001 & 0.4001 & 0.4 & 0.3 \\
$p_4$ & 0.5002 & 0.5002 & 0.5002 & 0.5 & 0.5 & -1.999 & -1.999 & -0.4 & -0.2 & 0.4001 & 0.4001 & 0.4 & 0.3 \\
$p_5$ & 0.5002 & 0.5002 & 0.5002 & 0.5 & 0.5 & -1.999 & -1.999 & -0.4 & -0.2 & 0.4001 & 0.4001 & 0.4 & 0.3 \\
$p_6$ & 0.5002 & 0.5002 & 0.5002 & 0.5 & 0.5 & -1.999 & -1.999 & -0.4 & -0.2 & 0.4001 & 0.4001 & 0.4 & 0.3 \\
$p_7$ & 0.5002 & 0.5002 & 0.5002 & 0.5 & 0.5 & -1.999 & -1.999 & -0.4 & -0.2 & 0.4001 & 0.4001 & 0.4 & 0.3 \\
$p_8$ & 0.5002 & 0.5002 & 0.5002 & 0.5 & 0.5 & -1.999 & -1.999 & -0.4 & -0.2 & 0.4001 & 0.4001 & 0.4 & 0.3 \\
$p_9$ & 0.5002 & 0.5002 & 0.5002 & 0.5 & 0.5 & -1.999 & -1.999 & -0.4 & -0.2 & 0.4001 & 0.4001 & 0.4 & 0.3 \\
$p_10$ & 0.5002 & 0.5002 & 0.5002 & 0.5 & 0.5 & -1.999 & -1.999 & -0.4 & -0.2 & 0.4001 & 0.4001 & 0.4 & 0.3 \\
$p_11$ & 0.5002 & 0.5002 & 0.5002 & 0.5 & 0.5 & -1.999 & -1.999 & -0.4 & -0.2 & 0.4001 & 0.4001 & 0.4 & 0.3 \\
$p_12$ & 0.5002 & 0.5002 & 0.5002 & 0.5 & 0.5 & -1.999 & -1.999 & -0.4 & -0.2 & 0.4001 & 0.4001 & 0.4 & 0.3 \\
\end{tabular}
\caption{Example positional assignment matrix. A solution must choose one entry from each row and each column, with the goal of mazimizing the total of chosen entries}
\label{Tab:PosAss}
\end{table} 

Solving the assignment problem tells $H_0$ how many of each slot it will have available for future players, but says nothing about how it will fill the flexible position slots, which can take players of multiple positions. To account for this, $H_0$ models separate flex share vectors $j_U$, $j_G$, and $j_F$ . They control the expected value of the fraction of flex spots that will be devoted to each position. If $H_0$ wants a mix of centers and power forwards, it can set its flex share vectors as 

$$
j_U = \begin{bmatrix}
C = 0.7 & PG = 0 & SG = 0 & PF = 0.3 & SF = 0
\end{bmatrix}
$$

$$
j_G = \begin{bmatrix}
PG = 0.3 & SG = 0.7
\end{bmatrix}
$$

$$
j_F = \begin{bmatrix}
PF = 1 & SF = 0
\end{bmatrix}
$$

This implies that the manager will fill its three utility spots with an expected value of 70\% centers and 30\% power forwards, etc. 

A vector $P$ is then calculated as the sums of players expected from each non-flex slots. Continuing the same example

\begin{itemize}
\item There is $1$ center slot left, plus $70\%$ of $3$ utility slots. So there are $3.1$ future Cs 
\item There is $1$ point guard left, plus $30\%$ percent of $2$ guard slots. So there are $1.6$ future PGs
\item There is $1$ shooting guard slot left, plus $70\%$ percent of $3$ utility slots. So there are $2.4$ future SGs 
\item $30\%$ percent of $3$ utility slots go to power forwards, plus $2$ forward slots. So there are $2.3$ future PFs
\item There is $1$ small forward slot remaining, so there is $1$ future SF
\end{itemize}

Altogether, 

$$
P = \begin{bmatrix}
C = 3.1 & PG = 1.6 & SG = 2.4 & PF = 2.9 & SF = 1
\end{bmatrix}
$$

Finally, the $P$ vector is combined with $\mu_C$ to get a positional adjustment $\mu_C P$. Combined with the effect of categorical weightings the expression for $X_\delta(j)$ is 

\begin{align*}
X_\delta(j) = & \left(N-K-1\right) * \Sigma * \left( v j_C^T - j_C v^T \right) * \Sigma * \frac{
   \left( - \gamma j_C - \omega v \right) \sqrt{\left(j_C -  \frac{v v^T \Sigma j_C}{v^T \Sigma v} \right) ^T \Sigma \left( j_C -  \frac{v v^T \Sigma j_C}{v^T \Sigma v}  \right) }
  }{j_C^T \Sigma j_C * v^T \Sigma v - \left( v^T \Sigma j_C \right) ^2}\\
  & + \mu_C P
\end{align*}

This specification for $X_\delta(j)$ is differentiable relative to $J_U$, $J_G$, and $J_C$. They have linear effects on $P$ so the derivative of e.g. the SG component of $J_G$ is just the $\mu_C$ row for SGs, multiplied by the number of guards decided by the assignment problem

\subsection{Modeling win probabilities $W$}

$H_0$ calculates the probability of victory for team $A$ based on the CDF of $X(j)$, which is a normal distribution, at zero. Calling its mean $\mu$ and the variance $\sigma^2$, the category victory probabilities are 

$$
w_c = \frac{1}{2} \left[ 1 + \erf  \left( \frac{\mu}{ \sqrt{2} * {\sigma}} \right) \right]
$$

The $c$ subscript is useful for keeping track of win probabilities across categories

\subsection{Modeling the objective function $V$} \label{Objective}

Once a $w_c$ is computed for each category, $H_0$ converts them into the objective function $V$. 

Note that the objective function is defined according to an arbitrary match-up. To get at the result for an arbitrary match-up, the objective functions are calculated for each opponent, then those results are averaged

\subsubsection{Each Category Objective}

The objective function for the Each Category format is 

$$
V(j) =  \sum_{c \in C} w_c(X(j))
$$

The gradient of this objective function is

$$
\nabla V(j) = \sum_{c \in C} PDF_c(X(j)) * \nabla X(j)
$$

Where PDF is the probability density function corresponding to the category differential distribution. Details of the calculation are included in Appendix \ref{apdx.ECGradient}

\subsubsection{Most Categories Objective}

The expression for probability of winning in Most Categories, assuming all categories are independent, is

\begin{align*}
V(j) = & \sum_{s \in S_W} \prod_{c \in C} f(s,c) * w_c(X(j)) + (1 - f(s,c))(1 - w_c(X(j))) \\
+ & \frac{1}{2} \sum_{s \in S_T} \prod_{c \in C} f(s,c) * w_c(X(j)) + (1 - f(s,c))(1 - w_c(X(j)))
\end{align*}

Where $S_W$ is a set of overall winning scenarios in terms of which individual categories are won and lost, and $S_T$ is a set of tying scenarios. $f(s,c_1)$ is one if the category $c_1$ is won in scenario $s$ and zero otherwise. 

The possibility of a category-level tie is irrelevant since category distributions are modeled as continuous variables, and therefore the probability of a tie is theoretically infinitesimal. However, overall ties have non-zero probability if the number of categories is even, which is why they are included in the objective function.

$H_0$ must consider each individual scenario in $S_W$, and if there are an even number of categories, $S_T$ as well. Fortunately, the number of winning scenarios in the typical 9-cat league is only 

$$
    \binom{9}{5} + \binom{9}{6}  + \binom{9}{7}  + \binom{9}{8} + \binom{9}{9}  = 256
$$

This is because there are $\binom{9}{5} $ scenarios where five categories are won and four are lost, $\binom{9}{6} $ scenarios where six categories are won and three are lost, etc. Manually checking each of these $256$ scenarios is tractable. Each winning scenario involves calculating $8$ multiplication steps so the total number of operations is no more than $2048$ per player. The efficiency of the operation can improved by computing probabilities with the procedure shown in Appendix \ref{apdx.MCTree}.

This objective function is differentiable. Details for how to calculate the gradient are included in Appendix \ref{apdx.MCGradient} and they result in

$$
\nabla V(j) = \left\{
\begin{array}{ll}
      \sum_{c_1 \in C} T(j, c_1)  *  PDF(X(j)) * \nabla X(j) & |C| \text{ is odd} \\[5pt]
      \frac{1}{2} \sum_{c_1 \in C} T(j, c_1)  *  PDF(X(j)) * \nabla X(j) & |C| \text{ is even}
\end{array} 
\right. 
$$

$T(j, c_1)$ represents the probability that a category is a ``tipping point'', that is, the probability that $c_1$ could be a deciding factor in the overall result. For 9-cat, it is defined as

$$
T(j, c_1) =  \sum_{s \in S_{c_1}(4,4)} \left( \prod_{c_2 \in C} f(s,c_2) * w_{c_2}(X(j)) + \left( 1 - f(s,c_2) \right) \left( 1 - w_{c_2}\left( X(j) \right) \right) \right)
$$

Where $S_{c_1}(n,m)$ is a set of scenarios across all categories except $c_1$, for which between $n$ and $m$ are wins. $S_{c_1}(4,4)$ is relevant in this case because $c_1$ could be the deciding category if four other categories are won and four are tied

\subsection{Optimizing in practice}

The value of $V(j)$ is $H_0$'s definition of success. Discovering the best value of $j$ to make $V(j)$ as high as possible requires optimization

\subsubsection{Assignment problem}

The positional model described in section \ref{Positions} requires the solution to an assignment problem. Fortunately, assignment problems are well-studied and efficient solutions are available. For the purpose of this paper a modified Jonker-Volgenant algorithm was used, as implemented by python's scikit-learn package (Scipy.org, 2016). 

An alternative to solving the assignment problem separately is explicitly modeling each decision variable and optimizing them for $V$. This would require an alternative optimization method besides gradient descent, because gradient descent only applies to problems with continuous variables

\subsubsection{Gradient descent}

With an almost-always differentiable $V(j)$ available, gradient descent can be performed. 

An important limitation to keep in mind is that $V(j)$ is not convex, because the cumulative distribution function of a normal distribution is not convex. This means that gradient descent will only find a local minima, rather than a global minima. 

The downside of only being able to optimize locally can be ameliorated with clever choice of initial conditions. Intuitively, it is reasonable to expect that the best strategy will be similar to the weights computed in the previous round, so $j$ is initialized as a mixture of default weights $v$ and the previously computed optimal weights. For the first round when there are no previous weights, the initial point for $j_C$ is $v$ perturbed with a factor of $\frac{1}{500}$ in the direction of the candidate players' expected statistics (using exactly $v$ leads to an undefined gradient). This way of doing gradient descent does not guarantee that an optimal point will be found, but it ensures that the local neighborhood in which gradient descent choices is a reasonable guess for the best solution.

Each round of gradient descent may alter the sums of $j_C$, $j_U$, $j_G$, and $j_F$. While scaling $j_C$ up or down has no effect on resultant players chosen, the parameters $\gamma$ and $\omega$ are easiest to calibrate when the scale of $j_C$ is held constant. For that reason $\gamma$ and $\omega$ are calibrated based on the sum of $j_C$ always being one, and $H_0$ re-calibrates all $j_C$s to sum to one after each step of gradient descent by dividing through by the sum. The same is done for $j_U$, $j_G$, and $j_F$, which need to sum to one by definition.

For the purposes of this paper, gradient descent was carried out with the Adam optimizer (Kingma, 2014)

\section{Simulation}

Simulated versions of NBA fantasy seasons, from 2004-05 to 2023-24, were run to provide reassurance that the logical foundations of $H_0$ are solid. 

The simulated seasons were twelve-team, thirteen-player head-to-head competitions. Each player's performance in each simulated season was generated by randomly sampling twenty weeks from their actual performances, excluding weeks for which they were injured. Only players with ten or more weeks of playing time were included. Teams were paired against each other weekly and winners were decided by which team had the most points by the end of the twenty week season. For these simulations, the player requirement was defined by the structure presented in section \ref{Positions}, with players eligible for positions that they were eligible for on Yahoo's fantasy basketball platform. 

Managers had access to weekly performance numbers for each relevant player, allowing them to calculate all relevant metrics. Z-scores based on the full league were used to choose $Q$, based on which managers calculated H-scores and G-scores. 

H-score was tested at seats zero, one, etc. while all other drafters were using G-score.  One thousand of the twenty-week simulated seasons were run for every draft seat, allowing for robust estimates of how well that strategy really would have performed in that situation with error bars no greater than by $ \frac{\sqrt{1000 * \frac{1}{2}{(1 - \frac{1}{2})}}}{1000} \approx 1.6\%$. These match-ups were run for both the Most Categories and Each Categories format. The H-score drafters used $\omega = 0.7$ and $\gamma = 0.25$ for parameters.

\begin{figure}[h]

\begin{subfigure}[h]{1\linewidth}
\includegraphics[scale = 0.4]{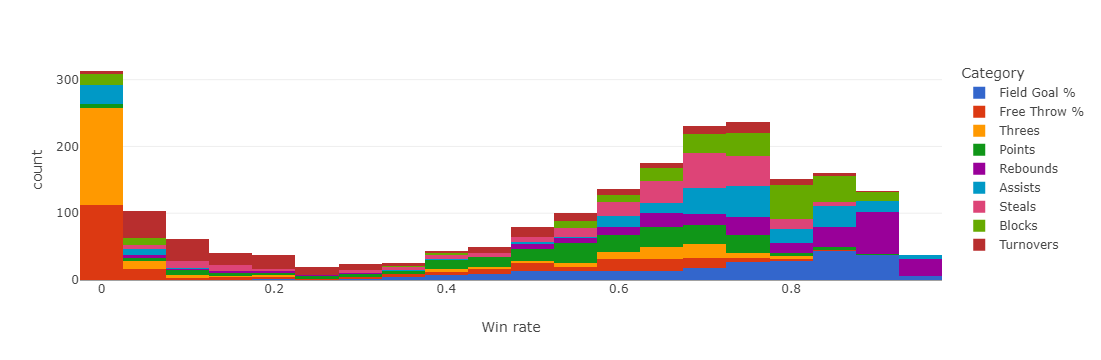}
\caption{Win rate histogram for Each Category}
\label{HistEC}
\end{subfigure}

\begin{subfigure}[h]{1\linewidth}
\includegraphics[scale = 0.4]{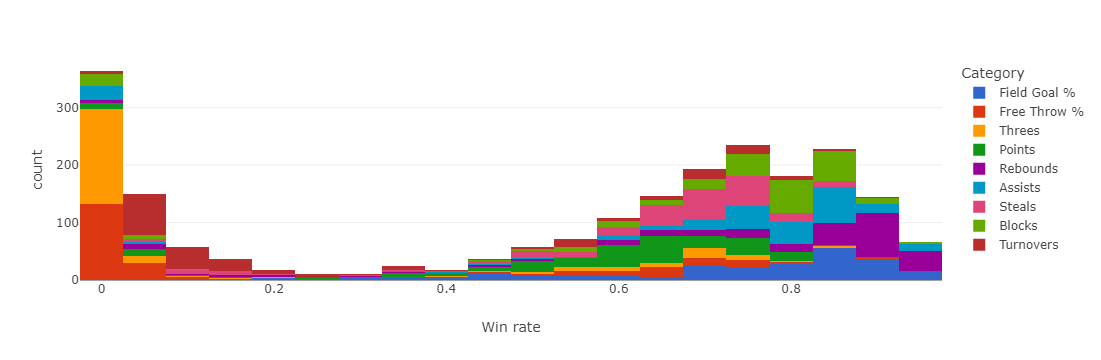}
\caption{Win rate histogram for Most Categories}
\label{HistMC}
\end{subfigure}

\caption{Histograms of win rate by category. Results are empirical, based on the simulations}
\label{Hist}

\end{figure}

\begin{figure}[h]
\includegraphics[scale = 0.39]{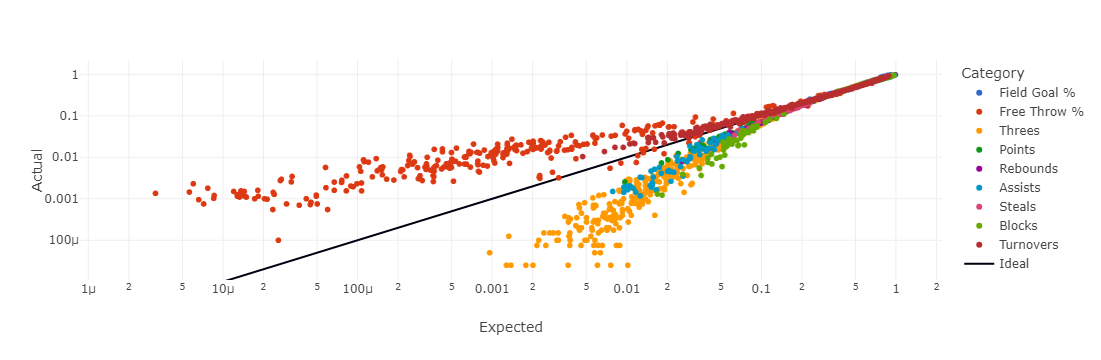}
\caption{Expected win rates against actual observed win rates, displayed in logarithmic scale. Results include both Each Categories and Most Categories}
\label{WinRates}
\end{figure}

\begin{figure}[h]
\begin{subfigure}[h]{1\linewidth}
\centering 
\includegraphics[scale = 0.4]{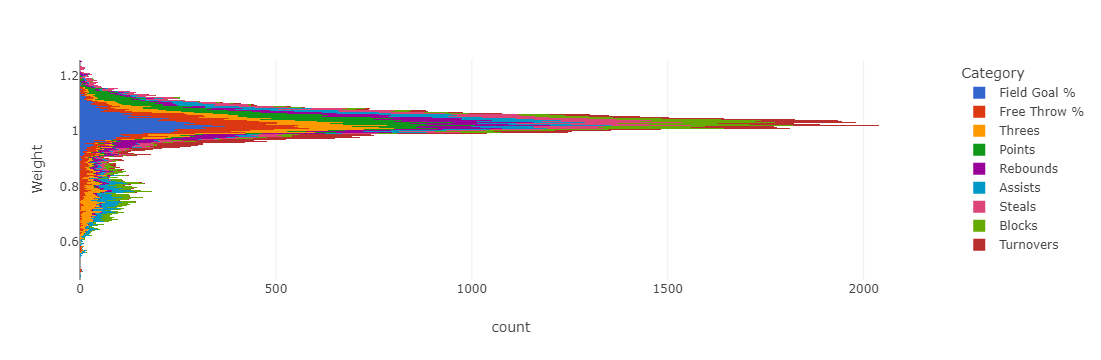}
\caption{Optimal first round pick weights for Each Category. 17\% of weights are below 0.95}
\label{WeightsEC}
\end{subfigure}

\begin{subfigure}[h]{1\linewidth}
\centering 
\includegraphics[scale = 0.4]{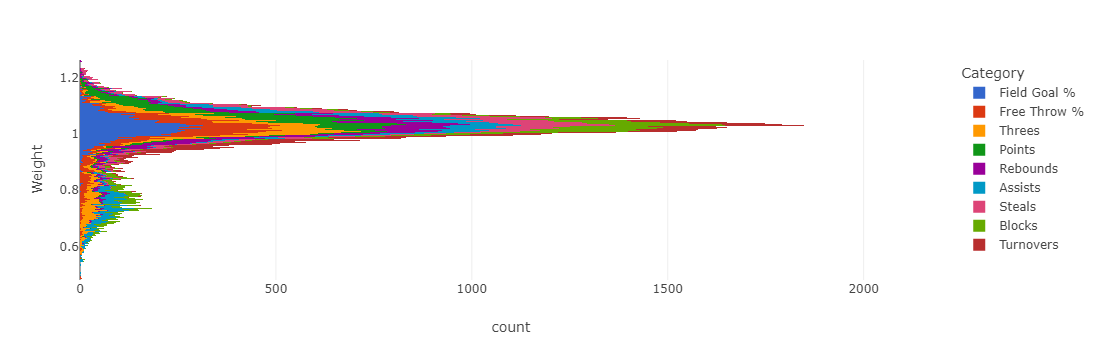}
\caption{Optimal first round pick weights for Most Categories. 18\% of weights are below 0.95}
\label{WeightsMC}
\end{subfigure}

\caption{Histogram of optimal first round pick weights. The top fifty candidates by G-score at each seat for each year were included. Weights are presented relative to default G-score weight}
\label{Weights}

\end{figure}

\begin{figure}[h]

\begin{subfigure}[h]{1\linewidth}
\centering 
\includegraphics[scale = 0.4]{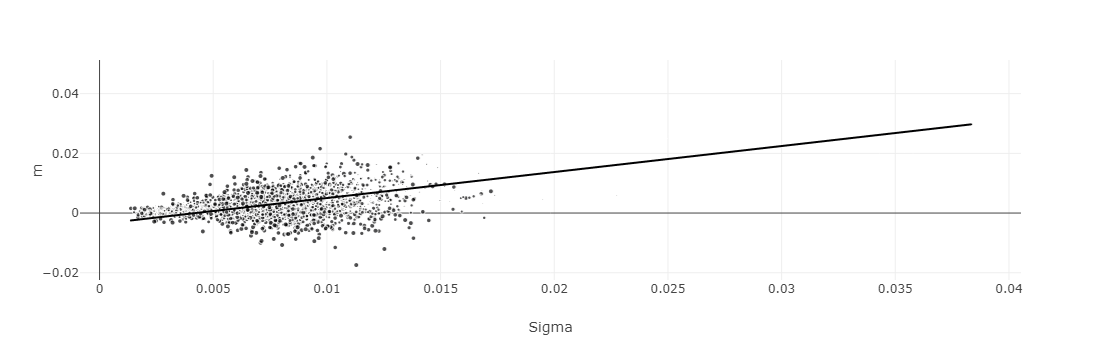}
\caption{Computed values of $\sigma$ vs eventual values of m, representing the $\omega$ parameter. The resultant best-fit line has a slope of $0.37$ with an R-square of $47\%$}
\label{Omega}
\end{subfigure}

\begin{subfigure}[h]{1\linewidth}
\centering 
\includegraphics[scale = 0.4]{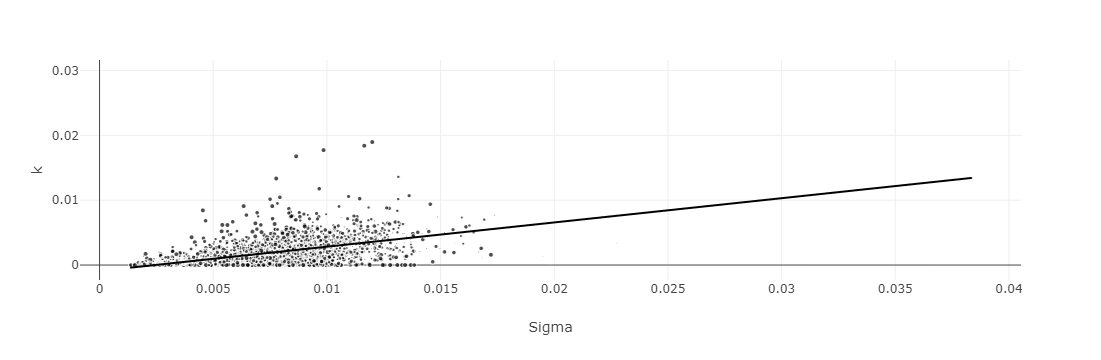}
\caption{Computed values of $\sigma$ vs eventual values of m, representing the $\gamma$ parameter. The resultant best-fit line has a slope of $0.87$ with an R-square of $46\%$}
\label{Gamma}

\end{subfigure}

\caption{Best-fit lines for calculating $\omega$ and $\gamma$, based on actual data from the simulations. Larger dots represent players drafted earlier, who have more data points for future picks}
\label{Calcs}

\end{figure}

The results are shown in Table \ref{HvG}, and Figures \ref{Hist}, \ref{WinRates}, \ref{Weights}, and \ref{Calcs}

\begin{table}[ht]
\centering 

\scalebox{0.8}{
\begin{tabular}{llrrrrrrrrrrrrr}
\toprule
 &  & 0 & 1 & 2 & 3 & 4 & 5 & 6 & 7 & 8 & 9 & 10 & 11 & Mean \\
\midrule
\multirow[t]{21}{*}{Each Category} & 2004-05 & 41.4\% & 20.4\% & 26.7\% & 27.4\% & 12.7\% & 13.3\% & 13.8\% & 20.6\% & 6.2\% & 3.7\% & 4.7\% & 5.3\% & 16.4\% \\
 & 2005-06 & 28.8\% & 12.0\% & 36.6\% & 42.5\% & 18.9\% & 19.7\% & 16.4\% & 14.8\% & 15.6\% & 18.5\% & 17.4\% & 20.2\% & 21.8\% \\
 & 2006-07 & 18.4\% & 20.1\% & 12.4\% & 15.2\% & 20.1\% & 9.2\% & 23.7\% & 5.9\% & 22.1\% & 14.0\% & 17.3\% & 12.9\% & 16.0\% \\
 & 2007-08 & 15.7\% & 8.6\% & 19.1\% & 35.1\% & 22.3\% & 7.2\% & 6.5\% & 4.9\% & 5.5\% & 9.0\% & 35.5\% & 29.7\% & 16.6\% \\
 & 2008-09 & 61.5\% & 65.7\% & 62.4\% & 8.5\% & 8.6\% & 12.0\% & 49.7\% & 49.0\% & 42.0\% & 6.7\% & 24.1\% & 7.8\% & 33.2\% \\
 & 2009-10 & 27.4\% & 29.8\% & 9.9\% & 40.6\% & 28.5\% & 34.1\% & 28.9\% & 24.9\% & 25.1\% & 6.3\% & 11.9\% & 8.6\% & 23.0\% \\
 & 2010-11 & 16.6\% & 18.6\% & 17.3\% & 28.5\% & 23.0\% & 25.1\% & 23.1\% & 25.6\% & 24.8\% & 30.9\% & 33.7\% & 17.9\% & 23.7\% \\
 & 2011-12 & 48.3\% & 39.9\% & 11.1\% & 19.4\% & 38.5\% & 37.8\% & 31.1\% & 22.4\% & 24.1\% & 26.9\% & 29.8\% & 27.7\% & 29.7\% \\
 & 2012-13 & 44.8\% & 27.7\% & 8.7\% & 8.8\% & 6.6\% & 9.2\% & 11.6\% & 12.8\% & 8.9\% & 17.5\% & 4.9\% & 5.1\% & 13.9\% \\
 & 2013-14 & 21.3\% & 7.4\% & 30.6\% & 21.6\% & 10.0\% & 29.9\% & 12.4\% & 4.6\% & 5.6\% & 6.0\% & 3.2\% & 9.3\% & 13.5\% \\
 & 2014-15 & 18.2\% & 37.7\% & 36.4\% & 16.9\% & 12.5\% & 14.5\% & 33.6\% & 32.8\% & 34.4\% & 31.9\% & 31.1\% & 9.2\% & 25.8\% \\
 & 2015-16 & 46.0\% & 24.4\% & 22.9\% & 11.6\% & 19.6\% & 8.9\% & 19.4\% & 22.3\% & 13.3\% & 30.1\% & 23.8\% & 24.6\% & 22.2\% \\
 & 2016-17 & 17.5\% & 16.7\% & 16.2\% & 27.1\% & 21.1\% & 29.2\% & 26.4\% & 10.8\% & 22.3\% & 15.8\% & 32.0\% & 18.9\% & 21.2\% \\
 & 2017-18 & 15.6\% & 16.7\% & 24.1\% & 34.8\% & 28.3\% & 30.7\% & 21.5\% & 32.0\% & 22.4\% & 15.2\% & 26.3\% & 30.3\% & 24.8\% \\
 & 2018-19 & 52.4\% & 20.9\% & 16.4\% & 27.5\% & 27.1\% & 26.6\% & 26.9\% & 26.7\% & 25.1\% & 28.3\% & 11.3\% & 13.5\% & 25.2\% \\
 & 2019-20 & 35.8\% & 17.1\% & 23.6\% & 33.8\% & 28.3\% & 24.2\% & 23.2\% & 32.6\% & 16.0\% & 14.8\% & 15.7\% & 6.3\% & 22.6\% \\
 & 2020-21 & 47.6\% & 12.8\% & 18.9\% & 18.9\% & 22.9\% & 40.1\% & 38.5\% & 38.3\% & 43.3\% & 41.0\% & 34.7\% & 34.5\% & 32.6\% \\
 & 2021-22 & 35.4\% & 23.6\% & 8.2\% & 5.7\% & 20.2\% & 23.0\% & 22.5\% & 12.1\% & 16.1\% & 13.7\% & 14.1\% & 7.3\% & 16.8\% \\
 & 2022-23 & 21.7\% & 9.3\% & 14.4\% & 21.9\% & 30.6\% & 33.5\% & 22.9\% & 14.7\% & 14.2\% & 10.1\% & 12.3\% & 11.9\% & 18.1\% \\
 & 2023-24 & 19.2\% & 26.5\% & 48.4\% & 8.9\% & 20.8\% & 10.4\% & 15.6\% & 16.2\% & 14.3\% & 12.3\% & 15.3\% & 10.6\% & 18.2\% \\
 & Mean & 31.7\% & 22.8\% & 23.2\% & 22.7\% & 21.0\% & 21.9\% & 23.4\% & 21.2\% & 20.1\% & 17.6\% & 20.0\% & 15.6\% & 21.8\% \\
\cline{1-15}
\multirow[t]{21}{*}{Most Categories} & 2004-05 & 51.1\% & 54.3\% & 49.0\% & 49.7\% & 26.5\% & 28.4\% & 21.1\% & 34.3\% & 9.1\% & 7.1\% & 11.5\% & 11.4\% & 29.5\% \\
 & 2005-06 & 23.7\% & 15.1\% & 56.2\% & 66.8\% & 59.3\% & 20.9\% & 32.5\% & 14.6\% & 11.2\% & 33.2\% & 33.5\% & 35.6\% & 33.5\% \\
 & 2006-07 & 23.2\% & 23.0\% & 12.9\% & 15.7\% & 17.8\% & 8.5\% & 9.2\% & 8.2\% & 23.9\% & 31.7\% & 34.4\% & 31.9\% & 20.0\% \\
 & 2007-08 & 24.4\% & 12.1\% & 15.3\% & 63.9\% & 59.8\% & 7.0\% & 11.0\% & 8.3\% & 11.2\% & 46.0\% & 45.8\% & 43.0\% & 29.0\% \\
 & 2008-09 & 76.8\% & 80.7\% & 80.5\% & 6.7\% & 7.0\% & 5.9\% & 7.8\% & 6.8\% & 9.5\% & 6.1\% & 48.7\% & 8.0\% & 28.7\% \\
 & 2009-10 & 66.2\% & 27.4\% & 45.8\% & 54.5\% & 52.7\% & 56.6\% & 56.2\% & 58.4\% & 59.9\% & 27.8\% & 28.3\% & 25.8\% & 46.6\% \\
 & 2010-11 & 52.1\% & 51.3\% & 51.2\% & 50.9\% & 44.7\% & 47.1\% & 44.1\% & 47.3\% & 45.5\% & 44.8\% & 50.1\% & 41.2\% & 47.5\% \\
 & 2011-12 & 71.1\% & 57.2\% & 53.7\% & 57.9\% & 55.0\% & 58.1\% & 58.7\% & 33.2\% & 35.1\% & 36.3\% & 60.2\% & 62.3\% & 53.2\% \\
 & 2012-13 & 40.1\% & 39.3\% & 15.2\% & 16.6\% & 15.1\% & 10.7\% & 23.8\% & 11.3\% & 13.7\% & 33.0\% & 18.6\% & 14.1\% & 20.9\% \\
 & 2013-14 & 35.8\% & 40.1\% & 39.8\% & 41.6\% & 17.8\% & 14.6\% & 14.4\% & 10.0\% & 10.0\% & 32.6\% & 10.0\% & 22.9\% & 24.1\% \\
 & 2014-15 & 66.4\% & 40.9\% & 29.3\% & 44.5\% & 38.1\% & 37.0\% & 53.3\% & 53.0\% & 51.1\% & 47.4\% & 49.1\% & 20.5\% & 44.2\% \\
 & 2015-16 & 44.6\% & 32.3\% & 36.4\% & 31.6\% & 30.7\% & 31.3\% & 24.9\% & 26.9\% & 26.4\% & 57.4\% & 52.0\% & 54.7\% & 37.4\% \\
 & 2016-17 & 39.6\% & 40.6\% & 38.0\% & 44.6\% & 58.6\% & 60.3\% & 65.2\% & 50.7\% & 47.5\% & 36.7\% & 61.3\% & 66.6\% & 50.8\% \\
 & 2017-18 & 23.7\% & 67.6\% & 65.6\% & 68.5\% & 52.6\% & 54.6\% & 49.7\% & 48.2\% & 35.9\% & 44.3\% & 48.5\% & 49.9\% & 50.7\% \\
 & 2018-19 & 60.8\% & 56.8\% & 48.1\% & 46.9\% & 63.7\% & 64.2\% & 43.7\% & 47.3\% & 37.6\% & 59.6\% & 34.0\% & 40.2\% & 50.2\% \\
 & 2019-20 & 45.0\% & 10.8\% & 8.9\% & 48.7\% & 48.3\% & 56.1\% & 54.7\% & 54.2\% & 36.9\% & 38.2\% & 36.4\% & 29.3\% & 38.9\% \\
 & 2020-21 & 50.6\% & 36.0\% & 37.2\% & 39.1\% & 40.3\% & 43.5\% & 53.3\% & 51.6\% & 57.7\% & 47.3\% & 56.7\% & 49.1\% & 46.9\% \\
 & 2021-22 & 53.1\% & 30.1\% & 12.5\% & 12.5\% & 53.9\% & 59.0\% & 57.3\% & 48.5\% & 59.3\% & 54.7\% & 50.0\% & 13.2\% & 42.0\% \\
 & 2022-23 & 11.8\% & 12.7\% & 12.6\% & 40.4\% & 39.7\% & 36.3\% & 40.1\% & 45.8\% & 46.0\% & 12.1\% & 44.7\% & 7.7\% & 29.2\% \\
 & 2023-24 & 62.8\% & 23.4\% & 55.2\% & 14.4\% & 13.3\% & 9.7\% & 27.7\% & 35.3\% & 31.1\% & 31.6\% & 32.0\% & 37.2\% & 31.1\% \\
 & Mean & 46.1\% & 37.6\% & 38.2\% & 40.8\% & 39.7\% & 35.5\% & 37.4\% & 34.7\% & 32.9\% & 36.4\% & 40.3\% & 33.2\% & 37.7\% \\
\cline{1-15}
\bottomrule
\end{tabular}
}

    \caption{Win rates for $H_0$ against a field of G-score drafters}
\label{HvG}
\end{table}

\section{Discussion}

\subsection{Simulation results}

\subsubsection{Performance}

Figure \ref{HvG} shows that $H_0$ performed well against a field of G-score agents. It won $21.8\%$ of its seasons in Each Category, and  $37.7\%$ of its seasons in Most Categories. Both marks are well above the baseline of random chance, which is $8.3\%$. 

It is not immediately obvious why $H_0$ performed better in Most Categories than Each Category. One possibility is that in Each Category, sub-optimal opponents can randomly string together a series of 9-0 victories, which can be difficult to overcome. A result for Most Categories is either 1-0 or 0-1, making it more stable

A few other observations can be made from Figure \ref{HvG}. One is that the algorithm generally performed better with higher draft seats. This tracks with the concept that the highest-value players have the most spread between each other, as is standard for many probability distributions including normal distributions. Top draft picks are very valuable and make the algorithm's job easier. 

Another observation is that the success of $H_0$ was not universal, especially in Each Category and with low draft picks. Its win percentage was as low as $3.2\%$ in 2013-2014 with the 11th pick in Each Category. This is not surprising, given both that the top drafters often have inbuilt advantages, and that the H-score algorithm uses a plethora of assumptions which are not correct

\subsubsection{Category win rates}

Figure \ref{Hist} demonstrates that $H_0$ implicitly understood the concept of punting. The two histograms show the win rate distributions for categories across every H-score drafter from the simulations. The bulk of the distributions' masses are centered slightly above the $50\%$ win rate, with a significant lower mode at $0\%$. In other words, the algorithm consistently over-performed in most of the categories, and hardly competed for the rest. 

Additionally, it can be seen that $H_0$ rarely invested so much into a category that it nearly guaranteed wins. The density at $100\%$ is quite low, especially compared to the density at $0\%$ which represents the opposite situation. This suggests that the H-scoring algorithm was successfully re-balancing by not overly focusing on categories that it was already strong in.  

The objective function gradients provide intuition on why $H_O$ learns how to punt and re-balance without being explicitly told to do so. The gradients are linearly related to category PDF values, which are thickest around zero, where the team is expected to win at a $50\%$ rate. This means that during gradient descent, the more average the team is performing in a category, the more that $H_0$ is trying to increase the weight for that category. A category that starts out average will be boosted to a bit above average. A category that is below average will have a low gradient, incentivizing $H_0$ to invest even less in that category, creating a snowball effect representing punting. And if a category is above average, $H_0$ will also invest less in it, keeping it slightly above $50\%$ without bringing it all the way to $100\%$.

Between Each Category and Most Categories, it is apparent that the distribution for Most Categories is more skewed to the extremes. It has a larger volume of categories around $0\%$, a thinner distribution around $20\%$ to $50\%$, and a higher center of mass on the right. This tracks with the idea that punting is most effective and worthwhile in Most Categories, because there is no marginal value in winning another category when the majority is already won. This intuition is borne out by the expression for the gradient; gradient magnitudes are proportional to the probability that a category ends up being a tipping point

\subsubsection{Predicting category win rates}

$H_0$ predicted win rates moderately well, though there was significant distortion on the lower end. Figure \ref{WinRates} shows expected win rates versus actual win rates on a logarithmic scale. Above $10\%$ or so, actual win rates match expected win rates closely. At lower probabilities, the algorithm over-predicted success for some categories (assists, three-pointers, and blocks) and under-predicted it for others (turnovers and free-throw percent). These distortions are likely due to incorrect assumptions made by the algorithm, particularly that all players contribute the same variance, and percentage statistics can be treated equivalently to counting statistics in the X-score basis

\subsubsection{Weights}

Figure \ref{Weights} shows optimal weights computed by the algorithm. Perhaps surprisingly, the algorithm did not bifurcate weights to an extreme degree between punting and not-punting. Instead, it took a more subtle ``soft-punting'' approach. It weighed most categories a bit above $100\%$ and compensated with a long tail below for punted categories, peaking around $75\%$ or so. The lower tail represented slightly $20\%$ of all category weights, representing one or two categories on average. 

Intuitively, one might expect that if a manager is punting, they should bring their weight for that category all the way to $0\%$. However, that is not necessarily optimal. Even if a manager has a very low chance of winning a category, that chance is never zero. And across a field of candidates with relatively similar overall value, even a small de-weighting of one category can significantly skew the expected statistics of the highest-value player. Bringing a weight all the way to $0\%$ might sacrifice the category more than is necessary in order to bolster performances in all of the other categories.

It should be remembered that the weights are estimates of best weights used for future players, with statistics distributed according to a simplified model. For player $p$ who the algorithm is choosing, the algorithm may implicitly be using a very different weighing mechanism, because it has actual statistics on available players and does not have to make guesses. 

It should also be kept in mind that the weights calculated by $H_0$ are subject to a series of assumptions and simplifications. including those discussed in Sections \ref{VarAss} and \ref{FormAss}. More work on these fronts could allow for more precise weights

\subsubsection{Turnovers}

Another observation from Figure \ref{Weights} which may be surprising is that the algorithm did not try to down-weight turnovers by default, which is a common tactic advocated for by fantasy analysts. Instead, turnover weights were distributed similarly to other categories. 

The oft-cited intuition behind down-weighting turnovers is that turnovers reward teams for having their players sit on the bench without touching the ball, which is the opposite of what managers want in general. That argument is not precisely logical, since scores for the other counting statistics naturally counterbalance the effect of turnovers. However, there is a way of re-framing the argument that cannot be accounted for by the logic behind $H_0$. As discussed in Section \ref{IndAss}, $H_0$ does not model the correlations between categories on the week-to-week performance level. It could be that for a team to be competitive in turnovers, it must be doing poorly overall, and therefore be unlikely to win overall. If an investment in turnovers is only useful when the week is lost already, then that investment has no value. 

The idea can be investigated by analyzing the gradient of the objective function $V(W)$ relative to strength in each category. With performances modeled as a correlated multivariate normals, the gradient of the objective can be estimated via simulation. With small advantage states added to each counting statistic (disadvantage for turnovers), the results are included in Table \ref{fig:Gradients}

It is apparent from these results that turnovers are roughly as important as other categories. As an advantage state is added, turnovers become less important, but other counting statistics also become less important at the same rate. 

This is reasonable upon careful consideration. If a match-up is close in terms of playing time, it is likely also close in terms of turnovers, making turnovers important. If one team has an advantage in playing time then it becomes unlikely that they will defy the odds and win turnovers, but at the same it also becomes unlikely for their opponents to flip the other counting statistics. Therefore turnovers retain their importance relative to other categories, except the percentage categories, which gain outsize importance because they are uncorrelated with playing time.

\begin{table}[h]
\small\addtolength{\tabcolsep}{-2pt}
\centering
\begin{tabular}{r r r r r r r r r r}
Victory probability & Points & Rebounds & Assists & Steals & Blocks & Threes & Turnovers & Free Throw \% & Field Goal \% \\ \midrule
& & & & & & & & & \\
            Most Categories & & & & & & & & &  \\ \midrule
50.0\% & 10.3\% & 6.8\% & 6.2\% & 9.0\% & 7.1\% & 6.6\% & 7.2\% & 7.1\% & 7.4\% \\ 
59.7\% & 10.0\% & 7.4\% & 6.7\% & 8.6\% & 5.9\% & 6.8\% & 7.0\% & 6.9\% & 7.1\% \\ 
68.9\% & 9.1\% & 6.4\% & 6.1\% & 8.0\% & 5.6\% & 6.0\% & 6.5\% & 6.4\% & 6.3\% \\ 
77.1\% & 8.4\% & 5.5\% & 5.1\% & 6.4\% & 4.6\% & 5.3\% & 5.2\% & 5.0\% & 5.5\% \\ 
83.9\% & 6.5\% & 4.6\% & 4.3\% & 5.3\% & 3.8\% & 4.1\% & 4.0\% & 4.3\% & 4.4\% \\ 
& & & & & & & & & \\
Each Category & & & & & & & & & \\ \midrule
    50.0\% & 34.0\% & 30.9\% & 27.7\% & 36.4\% & 30.6\% & 31.0\% & 33.4\% & 33.4\% & 34.4\% \\ 
54.3\% & 32.3\% & 30.9\% & 29.1\% & 36.4\% & 30.8\% & 30.3\% & 33.7\% & 33.4\% & 34.4\% \\ 
58.5\% & 29.6\% & 29.1\% & 27.3\% & 33.4\% & 28.2\% & 28.9\% & 30.9\% & 33.4\% & 34.4\% \\ 
62.3\% & 27.2\% & 27.0\% & 24.8\% & 29.2\% & 25.8\% & 26.0\% & 27.0\% & 33.4\% & 34.4\% \\ 
65.8\% & 22.8\% & 23.2\% & 21.7\% & 24.5\% & 23.0\% & 22.1\% & 24.2\% & 33.4\% & 34.4\% \\ 
\end{tabular}

\caption{Computed gradients with correlations included. Player performances are randomly sampled from a multivariate normal distribution, with correlations computed across players and performances. The advantage state is modeled with a small positive mean for the non-turnover counting statistics, and a negative one of the same magnitude for turnovers}
\label{fig:Gradients}
\end{table}

\subsubsection{Parameters}

With data from the simulations, it is possible to estimate $\omega$ and $\gamma$ using best-fit lines comparing $\sigma$ to actual values of $m$ and $k$. The results are in Figures \ref{Gamma} and \ref{Omega}. The computed values are reasonably close to the estimates used for the simulations, $0.25$ and $0.7$. 

It is also apparent from individual data points and R-squared values that the $\sigma$ is only so predictive of $m$ and $k$. Some of that is surely because of natural variation, but some of it could perhaps be reduced by more precise modeling

\subsection{Assumptions} \label{Ass}

\subsubsection{Teams must fit a certain position structure}

The idea that all teams must match an exact structure, and so long as they do all of their games count, is a simplification of real fantasy basketball. In reality positional structure is flexible; managers can bend on how balanced their team is based on how worthwhile it is for them to draft players of particular positions. Ideally H-scoring would implicitly understand this trade-off. However, it is difficult to quantify, since the degree to which a team has position-related issues depends on its exact players and how their schedules interact.

The position structure requirement is still helpful because it makes $H_0$ understand that it cannot load its team up entirely with players of the same position. Without understanding that, $H_0$ could make sub-optimal punting decisions. Also, it allows $H_0$ to see the value in players with statistics unusual for their position. Those players facilitate strong punting strategies because they leave many open slots for players that would fit the build later

\subsubsection{Performance distributions are known and do not change}

Player performances can drift over time for any number of reasons. This is not accounted for in $H_0$. 

One way in which this is problematic is that players don't always have the same number of games each week, leading to changes in expected weekly performances across seasons. Understanding how expectations differ from week to week, especially in light of particularly important weeks like playoff weeks, could improve the implementation of H-scoring. 

Another way in which this is problematic is that real fantasy basketball managers can ameliorate injury risk by swapping in un-injured players. This mitigates the risk of players prone to injuries, and makes them more value than their expected performances would indicate. Future work could perhaps build this logic into the algorithm.

Thirdly, it ignores the importance of general value in supporting flexibility. In real life, situations might change drastically, necessitating strategic pivots. Having high general value is useful for such situations, both since it increases the likelihood that a team will remain strong after changes, and because it gives the team more value for the trade market. $H_0$ doesn't see this value because it is certain that its model of how players will perform is correct, and does not understand the concept of trades. One way of incorporating this would be to use an ensemble of H-score and a more general metric like G-score, to pick players that are balanced between general value and value to the manager

\subsubsection{All players contribute the same level of variance} \label{VarAss}

It is convenient to assume that all players contribute the same amount of variance to a category, because player-level variance forecasts are generally not available. However this assumption is not entirely fair. It can be especially problematic in light of the idea that counting statistics are roughly Poisson variables, for which higher means imply higher variances. If a team is punting a category and has systematically low means, it likely has systematically low variance as well. Therefore $H_0$ will not estimate the category victory probability perfectly  

\subsubsection{Statistics for future picks follow a particular form} \label{FormAss}

$H_0$ makes liberal assumptions about the space of player statistics. Arguably, the assumptions do capture the main properties of the state space relevant to fantasy drafting. They specify that more valuable players are taken before less valuable players, and that there are trade-offs in weighting categories based on how they tend to correlate with each other. But they do this in a blunt way that is imprecise. In particular, there are two assumptions that are roundly unfounded and potentially problematic. 

The first is that other managers are choosing players in order according to their general value. In reality, opposing managers may be punting, and therefore preferring players with skewed statistical profiles. Or, they might have fundamentally different expectations of how players are expected to perform. There are almost innumerable reasons why a real manger might diverge from the simple model expected by $H_0$. 

The second potentially problematic assumption is that category-level statistics of relevant players are distributed as a multivariate normal. This is obviously not necessarily true. Player statistics can take on any kind of distribution, depending on the category, season, etc. For example it is well-known that blocks tend to have heavy right tails in general (Lloyd, 2023). 

Theoretically, it is possible to avoid needing assumptions about available players by modeling the dynamic fantasy basketball problem as a perfect-information sequential game. Every perfect-information sequential game has a subgame-perfect equilibrium which can be derived through backwards induction (Fudenberg, 1991). However, backwards induction requires evaluating every subgame outcome. In this case, evaluating every subgame outcome has a high state-space complexity which makes applying backwards induction practically difficult. 

Consider a snake draft. If there are $J$ available players, the first manager's initial pick breaks into $J$ subgames. Each of those players leads to $J-1$ subgames for the next manager, or $J *  \left( J-1 \right) $ in total. In general, the number of subgames at step $x$ of the draft is 

$$
\frac{J!}{ \left( J - x \right) !}
$$

If $K$ managers each choose $P$ players, then the total number of subgames is 

$$
\sum_{x = 1}^{K*P}\frac{J!}{ \left( J - x \right) !}
$$

Since all terms are positive, the final term with $x = K*P$ serves as a lower bound on the sum. With $J=500$ ( $\approx$ the number of NBA players) , $K=12$, and $P=13$ it is 

$$
\frac{500!}{ \left( 500 - 156 \right) !} \approx 10^{409}
$$

No modern computer is close to being able to evaluate that many subgames. Therefore, any practical advancement will still require some sort of heuristic. 

Future work could analyze the player statistics space with more sophistication, improving the Gaussian model to more accurately represent real data. This is likely a difficult task because the state space could look quite different depending on the overall player pool and which players have already been taken

\subsubsection{Percentage statistics can be treated equivalent to counting statistics}

For convenience, $H_0$ treats percentage statistics equivalent to counting statistics, once in the X-score basis. This is helpful because it allows counting statistics and percentage statistics to be modeled together as a multivariate normal distribution. The problem is that it is inaccurate, because it misses the effect of volume. Higher volume implies lower volatility, and this could be baked into a more sophisticated implementation of H-scoring 

\subsubsection{Category statistics are independent} \label{IndAss}

$H_0$ assumes that on a week-to-week basis, all categories are independent from each other. This is possibly quite untrue. If a team scores many threes on a particular week, they likely also scored many points. 

The reason for this omission is that incorporating week-to-week correlations would require using the CDF of a multivariate normal, for which there is no analytical expression (Genz, 2009). Player statistics could be more easily modeled as multivariate normals because only the means were required, not the CDF values. 

Handling this issue is not impossible, it is just computationally difficult. Perhaps a clever heuristic approach could make headway in the future. 

For what it is worth, week-to-week correlations were present in the simulations, and $H_0$ still performed well. So the omission is likely not extremely problematic 

\subsubsection{Managers want to to maximize expected value}

Real managers may care only about winning their leagues without a preference for last place versus fourth. This motivates an implementation of H-scoring which optimizes for upside potential instead of expected value. Also, real managers may want to focus most on playoff matches, since those are the most important matches for ultimate results. A future version of H-scoring could perhaps be tailored to this incentive structure as well

\subsubsection{Players up to $K+1$ from other teams are known}

This is another assumption of convenience, to make modeling easier. Future work could perhaps handle the case where the $K+1$'th player from another team is not known with more sophistication

\subsubsection{A local optimum is sufficient}

Non-global optimization is not ideal, but the trade-off is that non-convex optimization to find the actual global maximum would be more computationally intensive than gradient descent.

Another potential limitation is that the result will be non-robust. If the assumptions made by $H_0$ are quite wrong, and it needs to pivot its strategy in later rounds, there is no guarantee that it will be able to do so successfully since $H_0$ is only optimizing for the single optimal point. H-scoring might benefit from some kind of robust optimization, to improve resilience in face of inaccurate assumptions

\subsection{Rotisserie}

$H_0$ does not handle the Rotisserie format for computational reasons. If computation was not a limiting factor, it would be possible to estimate the probability of winning in Rotisserie with brute force, given approximate normal distributions for each teams' performances for each category. 
Unfortunately, the calculation is too intensive to be practical, even assuming that categories are uncorrelated. 

For a given category, there are $T!$ possible orderings, where $T$ is the number of teams. If $T$ is $12$ that translates to more than $479$ million. For each of those possible orderings, calculating the probability of the ordering occurring accurately would require intensive numerical integration. Further, the algorithm would have to analyze combinations of orders across categories.  With $9$ categories, the total number of orderings to analyze would be $1.32 * 10^{78}$. Both the integration and the combination steps would be too complex to be feasible with modern hardware. 

Future work could perhaps address Rotisserie with use of a clever heuristic

\section{Conclusion}

The H-scoring framework is introduced for dynamic optimization of draft picks. It is not as convenient to use as a static ranking list, but the described implementation $H_0$ does perform better, at least in simulations.

$H_0$ relies on many assumptions which can perhaps be ameliorated by better heuristics or eliminated by more involved mathematics. Improvements could be made in accounting for player-specific variance, more precisely modeling the distribution of future draft picks, adapting to choices of other managers, incorporating correlations between weekly category scores, strategizing around waiver wire moves, better modeling Rotisserie, and other areas.

\textit{Disclaimer: The views and opinions expressed in this article are those of the independent author and do not represent those of any organization, company or entity}

\begin{appendices}

\section{Adapting H-scoring to Auctions} \label{Auctions}

\subsection{Converting to auction value} \label{AuctionConv}

For drafting, the H-score calculation yields win probabilities based on which player is chosen. These are easy to use- the manager can just take the candidate player with the highest H-score.

However, the situation is not as simple for auctions. Auction managers need to quantify player values, not just rank them. Also, the raw H-score calculation yields a win probability for each candidate player if they could be selected without costing any money, which is not realistic. So some additional mechanisms are required to handle values for auctions. 

One way to equate H-scores to dollars is to subtract money (and corresponding value) from what the manager has remaining until they break even for taking the player. This is doable and theoretically works well, but is computationally expensive because it requires back-tracking through several calculations several times for each player.  

$H_0$ uses a less computationally intensive method, which is to start with a replacement player and various values of $X_m$ to see how level of cash affects H-scores. Approximate cash values can then be derived for players by comparing their H-scores to those of just adding cash, and finding the closest cash equivalents

\subsection{$H_0$ team decomposition for auctions} \label{AuctionDecomp}

Like in the snake draft context, with a certain number of players remaining, $H_0$ can assume some level of control over the weighting applied to those players to account for a punting strategy. The main difference is that $X(j)$ must be calculated in an auction-specific way

It is helpful to start by breaking down overall metrics in the following way
\begin{itemize}
\item $X_{A_\mu}  = X_s + X_p + X_r + X_m + X_{\delta}(j)$ where
    \begin{itemize}
  \item $X_s$ is the aggregate statistics of team $A$'s already selected players
  \item $X_p$ is the statistics of the candidate player
  \item $X_r$ is the statistics of aggregate statistics replacement-level players, filling all empty slots 
  \item $X_m$ is the general benefit of leveraging extra money to get above-replacement players
  \item $X_{\delta}(j)$ is the differential effect of punting strategy on the above-replacement players that will be selected instead of the replacement-level players. In essence, this is equivalent to $X_{\delta}(j)$ in the drafting context
  \end{itemize}
\item $X_{o_\mu} = X_{o_s} + X_{o_r} + X_{o_m}$ where
    \begin{itemize}
  \item $X_{o_s}$ is the aggregate statistics of team $O$'s already selected players
  \item $X_{o_r}$ is the aggregate statistics replacement-level players, filling all empty slots 
  \item $X_{o_m}$ is the general benefit of leveraging extra money to get above-replacement players
  \end{itemize}
\end{itemize}

Then 

$$
X(j) = X_{A_\mu}  - X_{o_\mu} =  X_s + X_p - X_{o_s} + X_r - X_{o_r} + X_m - X_{o_m} + X_{\delta} 
$$

This equation can be grouped into four parts 
\begin{itemize}
\item $X_s + X_p - X_{o_s}$: difference of known player statistics
\item $X_r - X_{o_r}$: difference of replacement-level values. E.g. if after adding the chosen player team $A$ has one more player selected already, then team $O$ has an additional replacement-level player which is subtracted out
\item $X_m - X_{o_m}$: the differential effect of team $A$ having more money remaining than team $O$
\item $X_{\delta}(j)$: differential as a result of punting strategy, defined in the same way as it was for the drafting context 
\end{itemize}

Take $M$ to be the number of extra players on team $A$ versus team $O$, including player $p$, and $R$ to be the statistics of a replacement-value player. Also take $L$ to be the amount of extra dollars team $A$ has, and $D$ to be the expected category benefit from one dollar worth of spending. The equation can then be rewritten to

$$
X(j) = X_s + X_p - X_{o_s} + MR + LD + X_{\delta}(j)
$$

$M$ and $L$ are readily available. $R$ and $D$ are harder to calculate 

Overall replacement value is easy to estimate with the highest G-score (the appropriate metric for static value) among players expected not to be drafted. $R$ is conceived of as an estimate of the statistical profile of a general player that could be picked up from the waiver wire or as a free agent, not necessarily mimicking the exact player seen to have the highest value. This necessitates careful handling of categories like turnovers, for which replacement-level players often are stronger than would otherwise be expected. $H_0$ distributes the overall replacement value evenly across players with a negative sign for turnovers. So with nine categories, it multiplies the replacement value (which is negative to start with) by $\frac{1}{7}$ for all categories except turnovers, and $- \frac{1}{7}$ for turnovers. It also divides by the $v$ vector to convert G-score value into X-score value.

$H_0$ estimates $D$ by taking the sum of available above-replacement value (weighted by the $v$ vector for generic value) over the sum of remaining money in the pool. To get per-category values, overall value is spread by per-category generic weight (as in, divided by the $v$ vector), with the value for turnovers inverted as it was for $R$.

This leaves $X_{\delta}(j)$ as the one remaining quantity to calculate, as it was for the snake draft context

\section{Estimating distribution of future picks} \label{apdx.xdel}

Define $x_{\delta_q}$ as the difference between candidate players' mean performances and category-level baselines. In order to model all category differential distributions as equivalent and smooth functions, it is useful to approximate $x_{\delta_q}$ as a correlated random Gaussian. While not necessarily accurate, this captures basic properties of the relationships between categories while facilitating a relatively straightforward approach to modeling. The covariance can be estimated empirically by making a matrix $X$, each row of which is $x_{\delta_q}$ of a player in $Q$, and calculating its covariance matrix.

If $x_{\delta_q}$ was all zeros, representing a baseline player, then its aggregate value in any weighting would also be zero. in terms of $j_C$, the chosen player $p$'s value will likely be above the value of a baseline player. It can be written that 

$$
j_C^T x_{\delta_p} = m
$$

On the other hand, in terms of generic value, the chosen player will likely be below baseline since the manager would need to sacrifice generic value in order to maximize value under $j_C$. Define 

$$
v = \sqrt{\frac{m_\tau^2 + m_\sigma^2}{m_\tau^2}}
$$

This translates from X-scores to G-scores, a measure of generic value. Then, 

$$
v^T x_{\delta_p} = - k
$$ 

The chosen player by construction has the highest $j_C^T x_{\delta_p}$ among players available. It is known that the expected value of the maximum of several normals with mean zero is roughly proportional to the standard deviation (Royston, 1982). That approximation can be invoked to declare that 

\begin{equation}
\label{eq1}
j_C^T x_{\delta_p} = \omega \sigma
\end{equation}

Where $\sigma$ is the standard deviation of $j_C^T x_{\delta_q}$ across relevant candidate players. 

It is reasonable to approximate the relationship between $m$ and $k$ as a linear function, because a more unique player will likely require searching through rankings for longer. It can then also be said that 

\begin{equation}
\label{eq2}
v^T x_{\delta_p} = - \gamma \sigma
\end{equation}

$\sigma$ takes some math to work out. Applying known linear algebra to the assumptions, $x_{\delta_q}$ has covariance matrix $A \Sigma A^T$ where $\Sigma$ is the covariance matrix describing $X$ and $A = I_9 - \frac{\Sigma v v^T}{v^T \Sigma v}$ (jlewk, 2022). So

$$
\sigma^2 = j_C^T A \Sigma A^T j_C
$$

Plugging in the definition of $A$ yields

$$
\sigma^2 = j_C^T \left( I_9 - \frac{\Sigma v v^T}{v^T \Sigma v} \right)  \Sigma \left( I_9 - \frac{\Sigma v v^T}{v ^T \Sigma v} \right)^T j_C
$$

Simplifying leads to

$$
\sigma^2 = \left(j_C -  \frac{v v^T \Sigma j_C}{v^T \Sigma v} \right) ^T \Sigma \left( j_C -  \frac{v v^T \Sigma j_C}{v^T \Sigma v}  \right) 
$$

$$
\sigma = \sqrt{\left(j_C -  \frac{v v^T \Sigma j_C}{v ^T \Sigma v} \right) ^T \Sigma \left( j_C -  \frac{v v^T \Sigma j_C}{v ^T \Sigma v}  \right) }
$$

Given this formulation, constraints \ref{eq1} and \ref{eq2}, and invoking the assumption that the underlying distribution of X is a multivariate normal, $x_\delta(j_C)$ can be derived. Applying more linear algebra, it is (jlewk, 2022)

$$
x_\delta(j_C) = \Sigma U^T \left( U \Sigma U^T \right)^{-1} b
$$

Where 

\[
U=
  \begin{bmatrix}
    & & & & v & & & & \\
    & & & & j & & & & 
  \end{bmatrix}
\]

\[
b=
  \begin{bmatrix}
    - \gamma \sqrt{\left(j_C -  \frac{v v^T \Sigma j_C}{v^T \Sigma (1)} \right) ^T \Sigma \left( j_C -  \frac{v v^T \Sigma j_C}{v^T \Sigma v}  \right) } \\
    \omega \sqrt{\left(j_C -  \frac{v v^T \Sigma j_C}{v^T \Sigma v} \right) ^T \Sigma \left( j_C -  \frac{v v^T \Sigma j_C}{v^T \Sigma v}  \right) }
  \end{bmatrix}
\]

This expression can be further simplified. Note that 

$$
U \Sigma U^T = 
  \begin{bmatrix}
  v^T \Sigma v & v^T \Sigma j_C \\
  v^T \Sigma j_C & j_C^T \Sigma j_C
\end{bmatrix}
$$

Making 

$$
\left( U \Sigma U^T \right)^{-1} = \frac{1}{j_C^T \Sigma j_C * v^T \Sigma v - \left( v^T \Sigma j_C \right) ^2}
  \begin{bmatrix}
  j_C^T \Sigma j_C & - v^T \Sigma j_C \\
  - v^T \Sigma j_C & v^T \Sigma v
\end{bmatrix}
$$

Since $U^T = \begin{bmatrix} v & j_C \end{bmatrix}$, 

$$
x_\delta(j_C) = \Sigma * \frac{\begin{bmatrix} v & j_C \end{bmatrix}   \begin{bmatrix}
  j_C^T \Sigma j_C & - v^T \Sigma j_C \\
  - v^T \Sigma j_C & v^T \Sigma v
\end{bmatrix} b }{j_C^T \Sigma j_C * v^T \Sigma v - \left( v^T \Sigma j_C \right) ^2}
$$

$$
= \Sigma * \frac{
  \begin{bmatrix}
  v * j_C^T \Sigma j_C -  j_C * v^T \Sigma j_C &
  - v * v^T \Sigma j_C + j_C * v^T \Sigma v 
\end{bmatrix} * b}{j_C^T \Sigma j_C * v^T \Sigma v - \left( v^T \Sigma j_C \right) ^2}
$$

Which further simplifies

$$
x_\delta(j_C) = \Sigma * \frac{
  \begin{bmatrix}
  \left( v j_C^T - j_C v^T \right) \Sigma j_C & 
  \left( j_C v^T - v j_C^T \right)  \Sigma v 
\end{bmatrix} * b}{j_C^T \Sigma j_C * v^T \Sigma v - \left( (\frac{1}{9})^T \Sigma j_C \right) ^2}
$$

$$
= \Sigma * \left( v j_C^T - j_C v^T \right) * \frac{
  \begin{bmatrix}
   \Sigma j_C & 
  -  \Sigma v
\end{bmatrix} * b}{j_C^T \Sigma j_C * v^T \Sigma v - \left( v^T \Sigma j_C \right) ^2}
$$

$$
= \Sigma * \left( v j_C^T - j_C v^T \right) * \frac{
   \left( - \Sigma j_C * \gamma  
  -  \Sigma v * \omega \right) * \sqrt{\left(j_C -  \frac{v v^T \Sigma j_C}{v^T \Sigma v} \right) ^T \Sigma \left( j_C -  \frac{v v^T \Sigma j_C}{v^T \Sigma v}  \right) } }{j_C^T \Sigma j_C * v^T \Sigma v - \left( v^T \Sigma j_C \right) ^2} 
$$

$$
= \Sigma * \left( v j_C^T - j_C v^T \right) * \Sigma * \frac{
   \left( - \gamma j_C - \omega v \right) \sqrt{\left(j_C -  \frac{v v^T \Sigma j_C}{v^T \Sigma v} \right) ^T \Sigma \left( j_C -  \frac{v v^T \Sigma j_C}{v^T \Sigma v}  \right) }
  }{j_C^T \Sigma j_C * v^T \Sigma v - \left( v^T \Sigma j_C \right) ^2}
$$

Finally, this value must be multiplied by the number of picks remaining. So the result is 

$$
X_\delta(j_C) =  \left(N-K-1\right) * \Sigma * \left( v j_C^T - j_C v^T \right) * \Sigma * \frac{
   \left( - \gamma j_C - \omega v \right) \sqrt{\left(j_C -  \frac{v v^T \Sigma j_C}{v^T \Sigma v} \right) ^T \Sigma \left( j_C -  \frac{v v^T \Sigma j_C}{v^T \Sigma v}  \right) }
  }{j_C^T \Sigma j_C * v^T \Sigma v - \left( v^T \Sigma j_C \right) ^2}
$$

\section{Gradient of H-score for Each Category}\label{apdx.ECGradient}

The gradient of the objective function with respect to $j$ is 

$$
\nabla V(j) = \sum_{c \in C} \nabla w_c(X(j))
$$

By the chain rule

$$
\nabla V(j) = \sum_{c \in C} \nabla_x w_{c}(X(j)) * \nabla X(j)
$$

Because $w_c$ is the cumulative distribution function of a normal distribution, this can be rewritten to

$$
\nabla V(j) = \sum_{c \in C} PDF_c(X(j)) * \nabla X(j)
$$

\section{Efficient computation for Most Categories}\label{apdx.MCTree}

The probability of match-up victory can be more efficiently computed using a tree. The top layer is winning points vs losing points, the next layer is winning points/winning rebounds vs winning points/losing rebounds etc. Each node stores the probability of the scenario occurring, and the children can be computed with one multiplication step each. Any node that represents five or more losses can be pruned. At layer $6$ for example, there are $\binom{6}{5}$ nodes representing $5$ losses and one win, and $\binom{6}{6} $ nodes representing $6$ losses, all of which can be ignored. The total nodes requiring multiplication in each layer, starting from the second layer, are shown in Table \ref{fig:CombCalc}.

\begin{table}[ht]
\centering
\begin{center}
\begin{tabular}{ r c }
Layer & Operations \\ \midrule
2 & $2^2 = 4$ \\
3 & $2^3 = 8$ \\
4 & $2^4 = 16$ \\
5 & $2^5- \binom{5}{5} = 31$ \\
6 & $2^6 - \binom{6}{5}  - \binom{6}{6}  = 57$ \\
7 & $2^7 - \binom{7}{5} - \binom{7}{6}  - \binom{7}{7}  = 99$ \\
8 & $2^8 - \binom{8}{5} - \binom{8}{6}  - \binom{8}{7}  - \binom{8}{8}  = 163$ \\
9 & $256$ as calculated earlier \\
\end{tabular}
\end{center}
\caption{Calculation of total number of scenarios}
\label{fig:CombCalc}
\end{table}

The total number of calculations add up to $634$, for a $69\%$ reduction in multiplication operations. This produces a relatively tractable operation, albeit a complicated one, to calculate the predicted win probability between two teams

\section{Gradient of H-score for Most Categories}\label{apdx.MCGradient}

The objective function has already been described as 

\begin{align*}
V(j) = & \sum_{s \in S_W} \prod_{c \in C} f(s,c) * w_c(X(j)) + (1 - f(s,c))(1 - w_c(X(j))) \\
+ & \frac{1}{2} \sum_{s \in S_T} \prod_{c \in C} f(s,c) * w_c(X(j)) + (1 - f(s,c))(1 - w_c(X(j)))
\end{align*}

Where $S_W$ is a set of scenarios for which the relevant player wins, $S_T$ is the same for ties, and $f(c,s)$ is a binary equalling $1$ if category $c$ is won in scenario $s$ and $0$ otherwise.

The case when there are an odd number of categories, and therefore no ties, is simplest. So assume that $S_T$ is empty for now. 

Because $\nabla \prod_i x_i = \sum_i \nabla x_i * \prod_{l, \neq i} x_l$, it can be said that

\begin{align*}
\nabla V(j) = \sum_{s \in S} \sum_{c_1 \in C} \left( f(s,c_1) * \nabla w_{c_1}(X(j)) -  \left( 1 - f(s,c_1) \right) \nabla w_{c_1}(X(j)) \right) \\
\left( \prod_{c_2 \in C} (f(s,c_2) * w_{c_2})(X(j)) + \left( 1 - f(s,c_2) \right) \left( 1 - w_{c_2}(X(j)) \right) \right) 
\end{align*}

Or, rearranging terms, 

\begin{align*}
\nabla V(x) = \sum_{s \in S} \sum_{c_1 \in C} \left( f(s,c_1) * \nabla w_{c_1} * (\prod_{c_2 \in C} f(s,c) * w_{c_2}(X(j)) + \left( 1 - f(s,c_2) \right) \left( 1 - w_{c_2}(X(j)) \right) \right) \\
- \left( (1 - f(s,c_1)) \nabla w_{c_1}(X(j)) * (\prod_{c_2 \in C} \left( f(s,c_2) * w_{c_2}(X(j)) + \left( 1 - f(s,c_2) \right) \left( 1 - w_{c_2}(X(j)) \right)   \right) \right) 
\end{align*}

The sum terms are interchangeable so 

\begin{align*}
\nabla V(j) = \sum_{c_1 \in C} \Biggl( \sum_{s \in S}  \left( f(s,c_1) * \nabla w_c{c_1} * (\prod_{c_2 \in C} f(s,c_2) * w_{c_2}(X(j)) + \left( 1 - f(s,c_2) \right) \left( 1 - w_{c_2}(X(j)) \right) \right) \\
- \sum_{s \in S} \left( (1 - f(s,c_1)) \nabla w_{c_1}(X(j)) * (\prod_{c_2 \in C} \left( f(s,c_2) * w_{c_2}(X(j)) + \left( 1 - f(s,c_2) \right) \left( 1 - w_{c_2}(X(j)) \right)   \right) \right) \Biggr)
\end{align*}

Now it is useful to define some additional scenario sets. $S_{c_1}(n,m)$ is a set of scenarios across all categories except $c_1$, for which between $n$ and $m$ are wins. Since $f(s,c_1)$ is 1 if and only if category $c_1$ is a win, and S consists of sets with $\frac{|C| + 1}{2}$ or more wins, the first product term is relevant for scenarios with $\frac{|C| - 1}{2}$ or more wins among the categories that are not $c_1$. Since $1- f(s,c_1)$ is 1 if and only if category $c_1$ is a loss, the second term is relevant for scenarios with $\frac{|C| + 1}{2}$ or more wins among the other categories. Also, it is helpful to shorten $\frac{|C| - 1}{2}$ as $n$. Subbing in the new notation

\begin{align*}
\nabla V(x) = \sum_{c_1 \in C} \Biggl( \sum_{s \in S_{c_1}(n,|C|)}  \left( \nabla w_{c_1} * \prod_{c_2 \in C} f(s,c_2) * w_{c_2}(X(j)) + \left( 1 - f(s,c_2) \right) \left( 1 - w_{c_2}(X(j)) \right) \right) \\
- \sum_{s \in S_{c_1}(n+1,|C|)} \left( \nabla w_{c_1}(X(j)) * (\prod_{c_2 \in C} \left( f(s,c_2) * w_{c_2}(X(j)) + \left( 1 - f(s,c_2) \right) \left( 1 - w_{c_2}(X(j)) \right)   \right) \right) \Biggr)
\end{align*}

The terms from scenarios with $n + 1$ or more wins cancel, since they are on both sides of the subtraction. 

$$
\nabla V(j) = \sum_{c_1 \in C} \sum_{s \in S_{c_1}(n,n}  \left( \nabla w_{c_1}(X(j)) * (\prod_{c_2 \in C} f(s,c_2) * w_{c_2}(X(j)) + \left( 1 - f(s,c_2) \right) \left( 1 - w_{c_2} \right) \right)
$$

Or 

$$
\nabla V(j) = \sum_{c_1 \in C} \nabla w_{c_1}(X(j))  \left( \sum_{s \in S_{c_1}(n,n)} (\prod_{c_2 \in C} f(s,c_2) * w_{c_2}(X(j)) + \left( 1 - f(s,c_2) \right) \left( 1 - w_{c_2}(X(j)) \right) \right)
$$

The large expression in parentheses can be thought of as a ``tipping point'' probability. For a given $c_1$, it is the probability that the other categories include exactly $n$ wins and $n$ losses, making $c_1$ the decisive category. It is intuitively logical that the tipping point probability would be a multiplier of the overall gradient, since it provides the exact probability that the category is relevant to winning chances.  

Defining 

$$
T(j, c_1) =  \sum_{s \in S_{c_1}(n,n)} \left( \prod_{c_2 \in C} f(s,c_2) * w_{c_2}(X(j)) + \left( 1 - f(s,c_2) \right) \left( 1 - w_{c_2}\left( X(j) \right) \right) \right)
$$

Yields 

$$
\nabla V(j) = \sum_{c_1 \in C} T(j, c_1)  * \nabla w_{c_1}(X(j)) 
$$

Or 

$$
\nabla V(j) = \sum_{c_1 \in C} T(j, c_1)  *  PDF(X(j)) * \nabla X(j)
$$

This is the same as the gradient for each-category, just with the extra $T(j, c_1)$ term.

The definition above of $\nabla V(j)$ was designed only for the case when the number of categories was odd, and therefore ties were impossible. Fortunately, it can be easily extended to the even case. 

In the even case, tipping points can change the result by one half-step by flipping a loss to a tie, a tie to a loss, a win to a tie, or a tie to a win. There are more tipping points in a sense, because any tie scenario is a tipping point, but the influence of a tipping point is half as strong. So the resulting gradient is 

$$
\nabla V(j) = \frac{1}{2} \sum_{c_1 \in C} T(j, c_1)  *  PDF(X(j)) * \nabla X(j)
$$

\end{appendices}

\bibliographystyle{agsm}

\end{document}